\documentclass{svproc}
\usepackage{amsmath,bm,bbm,latexsym,soul,nicefrac,young,youngtab}
\usepackage{graphicx,subfigure,multirow,dcolumn}

\usepackage{url}

\newcommand{\be}{\begin{eqnarray}}
\newcommand{\ee}{\end{eqnarray}}

\begin{document}
\mainmatter              
\title{Testing the Kerr nature with binary black hole inspirals}
\titlerunning{Testing Kerr nature with BBH inspirals}  
%
\author{Swarnim Shashank\inst{1} \and Cosimo Bambi\inst{1,2} \and Rittick Roy \inst{3}}
\authorrunning{S. Shashank et al.} 
%
\tocauthor{Swarnim Shashank, Cosimo Bambi, and Rittick Roy}
\institute{Center for Astronomy and Astrophysics, Center for Field Theory and Particle Physics and Department of Physics, Fudan University, Shanghai 200438, People's Republic of China
\and
School of Humanities and Natural Sciences, New Uzbekistan University, Tashkent 100001, Uzbekistan
\and
Anton Pannekoek Institute for Astronomy, University of Amsterdam, 1098 XH, Amsterdam, The Netherlands}

\maketitle              

\begin{abstract}
The theory of general relativity (GR) is the standard framework for the description of gravitation and the geometric structure of spacetime. With the recent advancement of observational instruments, it has become possible to probe the strong field regime to test GR. We present the constraints obtained from the binary black hole inspiral data of the LIGO-Virgo-Kagra (LVK) gravitational wave (GW) observations on the deformations of some popular parametrized non-Kerr metrics.
\end{abstract}
\section{Introduction}
Shortly after formulating general relativity (GR), Einstein derived the emission of gravitational radiation \cite{Einstein:1916cc}, now popularly termed gravitational waves (GWs). After 100 years, in 2015, the first detection of GWs from the coalescence of two black holes (BHs) was made by the LIGO labs \cite{first_gw_bbh_2016}, and since then, around 100 events have been observed \cite{GWTC1,GWTC2,GWTC3}.
Using the binary BH data observed by the LIGO-Virgo-Kagra (LVK) collaboration, tests of GR can be performed \cite{GRtest1,GRtest2,GRtest3}. One such test is the test of the Kerr hypothesis. The Kerr hypothesis states that astrophysical black holes are Kerr in nature \cite{Kerr:1963ud}. Testing the Kerr nature of a black hole is usually performed by assuming a parametrized spacetime where, in addition to the mass and spin, there are extra non-Kerr parameters that show a deviation from the Kerr geometry (see, e.g., \cite{Bambi:2015kza,ghasemi2016note,konoplya2016,mazza2021novel,Riaz:2022rlx,shashank2022,Shashank:2023erf,Johannsen:2013szh}). Recent efforts have focused on testing the Kerr hypothesis using electromagnetic spectrum data from X-ray and radio observations \cite{Bambi:2021chr,Tripathi:2020yts,EventHorizonTelescope:2022xqj,psaltis2020gravitational} and GW observations \cite{shashank2022,Riaz:2022rlx,Shashank:2023erf,cardenas2020gravitational}.

This contribution is based on Refs.~\cite{shashank2022,Riaz:2022rlx}, where GW data from the events observed by the LVK are used to constrain the non-Kerr parameters of some popular metrics. In the rest of the article, we dedicate each section to one metric. We very briefly explain the metric and its deformation parameters and then present the constraints obtained from the LVK data.

\section{KRZ metric}
The Konoplya, Rezzolla \& Zhidenko (KRZ) metric \cite{konoplya2016} is one of the popular parametrized axisymmetric metrics studied in the literature. In our methods, we study the inspirals of the GW sources; hence, we only consider equatorial circular orbits ($\theta = \pi/2$ and $\dot{\theta} = 0$). We set no spin ($a_* = 0$) because the spin effects on the deformation only enter at higher orders. In such a limit, the line element of the KRZ metric reads
\be
    d s^{2}&=&-\frac{N^{2}-W^{2}}{K^{2}} ~d t^{2}-2 W r ~d t d \phi
    +K^{2} r^{2} ~d \phi^{2} + \Sigma r^2 d\theta^{2} + \frac{\Sigma B^{2}}{N^{2}} ~  d r^{2} \, , 
\label{eq:krz_metric_equatorial}
\ee
where
\be
    N^2 &=& \left( 1 - \frac{2M}{r} \right)~\left[ 1 + \frac{8 M^3 \delta_1}{r^3} \right] \, , \quad
    B = 1 + \frac{4 M^2 \delta_4}{r^2} \, , \\
    \Sigma &=& 1, \quad K^2 = 1 \, , \quad
    W = \frac{8 M^3 \delta_2}{r^3} \, .
\label{eq:coeff}
\ee
The full metric without these assumptions can be seen in Ref.~\cite{konoplya2016,shashank2022}. From the metric, we follow the parametrized post-Einsteinian (ppE) formalism \cite{Yunes_Pretorius_2009} to derive the parametrized GW phase, which can then be used to test the Kerr nature by constraining the deformations using the LVK data. In our analysis, we only recover the $\delta_1$ and $\delta_2$ parameters \cite{shashank2022}.

\subsection{$\delta_1$}
From the normalization condition of the four-velocity, $u^{\mu} u_{\mu} = -1$,
\be
g_{rr} \dot{r}^2 = -1 - g_{tt} \dot{t}^2 - g_{\phi\phi} \dot{\phi}^2 \equiv V_{\mathrm{eff}} \, ,
\ee
introducing the effective potential $V_{\mathrm{eff}}$,
\be
\label{eq:Veff_full_delta1}
    V_{\mathrm{eff}} &=& -1 + E^2 + \frac{2 M}{r} + \frac{L^2 (2 M-r)}{r^3} + \frac{8 \delta_1 M^3 (2 M-r)}{r^4} \nonumber\\ 
    && + \frac{8 \delta_1 L^2 M^3 (2 M-r)}{r^6} +\mathcal{O}[{\delta_1}^{2}] \, .
\ee
where $E$ and $L$ are the specific energy and the specific angular momentum of a test-particle in circular orbit. They can be calculated by the condition $V_{\mathrm{eff}} = dV_{\mathrm{eff}}/dr = 0$. In the far-field limit, $L = r^2 \dot{\phi} \rightarrow r^2 \Omega$, where $\Omega = d\phi/dt$ is the angular velocity of the body as measured by a distant observer, we get a modified Kepler's law
\be \label{eq:mod_kep_delta1}
    \Omega^2 = \frac{M}{r^3} \left[ 1 + \frac{3 M}{r} + \frac{9 M^2}{r^2} - \frac{12 M^2}{r^2}\delta_1 + \mathcal{O} \left({\delta_1}^{2}, \frac{M^3}{r^3} \right) \right] \, . \nonumber\\
\ee
The power of $M$ in the $\delta_1$ term represents the PN order \cite{Carson_Yagi_2020_beyond-kerr}, so we can say that $\delta_1$ enters at 2PN. The orbital phase can be found by,
\be
    \phi(\nu) = \int^{\nu} \Omega ~dt 
    = \phi_{\mathrm{GR}}(\nu)-\frac{25}{4 \eta}(2 \pi m \nu)^{-1 / 3} \delta_1+\mathcal{O}\left[{\delta_1}^{2},(2 \pi m \nu)^{0}\right] \, ,
\ee
Taking the Fourier transform,
\be\label{eq:GW_phase_delta1}
    \Psi_{\mathrm{GW}}(f)=\Psi_{\mathrm{GW}}^{\mathrm{GR}}(f)-\frac{75}{8} u^{-1/3} \eta^{-4/5} \delta_1 + \mathcal{O}[{\delta_1}^{2}, u^0] \, , \nonumber\\
\ee
where $u = \eta^{3/5} \pi m f$. Comparing Eq.~\ref{eq:GW_phase_delta1} to the ppE framework \cite{Yunes_Pretorius_2009} and the LVK parametrization \cite{Yunes_Yagi_Pretorius_2016,lalsuite},
\be
\label{eq:beta_ppe}
    \beta = -\frac{75}{8} \eta^{-4/5} \delta_1 =  \frac{3}{128} \varphi_{4} \delta \varphi_{4} \eta^{-4/5}\, .
\ee
where, $\varphi_{4}$ is the PN phase at 2PN and has the form and  $\delta \varphi_4$ is the deviation from GR \cite{GRtest1,GRtest2,GRtest3} given as $\varphi_i \rightarrow (1 + \delta \varphi_i) \varphi_i$ ($i$ represents different PN phase). Finally, we find
\be\label{eq:d1-p4}
   \delta_1 = -\frac{1}{400} \varphi_{4} \delta \varphi_{4} \, .
\ee
which can be fit using the posteriors released by LVK. We analyze all the events present in the GWTC-2 catalogue and find a combined constraint by assuming each distribution follows a Gaussian distribution. For $\delta_1$ the constraint obtained are,
$$
\delta_1 = \begin{cases} -0.02 \pm 0.04 \quad &\text{({\tt IMRPhenomPv2})} \\ 0.02 \pm 0.05 \quad &\text{({\tt SEOBNRv4P})} \end{cases}.
$$
{\tt IMRPhenomPv2} and {\tt SEOBNRv4P} are the two waveform models used for the analysis by LVK. The constraints on each individual GWTC-2 events can be found in Ref.~\cite{shashank2022}.

\subsection{$\delta_2$}
In the case of $\delta_2$ deformation parameter, one obtains,
\be
\label{eq:mod_kep_delta2}
    \Omega^2 = \frac{M}{r^3}\left[ 1 + \frac{3 M}{r} + \frac{9 M^2}{r^2} + \frac{64 M^{5/2}}{r^{5/2}}\delta_2 + \mathcal{O} \left({\delta_2}^{2}, \frac{M^3}{r^3} \right) \right] \, ,\nonumber\\
\ee
and
\be\label{eq:d2-p5}
    \delta_2 = -\left(\frac{7729 \pi}{182784}-\frac{13 \pi \eta}{2176} \right) \delta \varphi_5 \, .
\ee
where $\varphi_5$ and $\delta \varphi_5$ are the phase and phase correction for 2.5PN. The combined constraints obtained on $\delta_2$ are,
$$
\delta_2 = \begin{cases} 0.01 \pm 0.02 \quad &\text{({\tt IMRPhenomPv2})} \\ 0.00 \pm 0.02 \quad &\text{({\tt SEOBNRv4P})} \end{cases}.
$$
The constraints on individual GWTC-2 events can be found in Ref.~\cite{shashank2022}.

\section{Johannsen metric}
The Johannsen metric \cite{Johannsen:2013szh} is another popular axisymmetric parametrized metric for measuring deviations from Kerr spacetime. For equatorial geodesics, the metric reads,
\be
    \label{eq:johannsen_metric}
	d s^2 = -\frac{\Sigma \Delta}{B^2} ~d t^{2} + \frac{\Sigma}{\Delta} ~d r^{2}
	+ \Sigma \left( d \theta^{2} + d \phi^{2} \right) \, ,
\ee
where
\be
    \label{eq:johannsen_coeff}
	 \Sigma = r^2 \, , \quad \Delta = r^2 - 2Mr \, , \quad B = r^2 A \, , \quad
	 A = 1 + \alpha_{13} ~\left(\frac{M}{r}\right)^3 \, .
\ee
Here, only one of the deformation parameters ($\alpha_{13}$) of the metric is considered. The full metric with all other deformation parameters can be found in Ref.~\cite{Johannsen:2013szh}. From here we obtain,
\be \label{eq:mod_kep_alpha13}
    \Omega^2 = \frac{M}{r^3} \left[ 1 + \frac{3 M}{r} + \frac{9 M^2}{r^2} - \frac{3 M^2 \alpha_{13}}{r^2} + \mathcal{O} \left({\alpha_{13}}^{2}, \frac{M^3}{r^3} \right) \right] \, . \nonumber\\
\ee
and similar to $\delta_1$ of the KRZ metric in the previous section we get a 2PN correction,
\be
   \alpha_{13} = -\frac{1}{100} \varphi_{4} \delta \varphi_{4} \, ,
\ee
The combined constraints obtained for $\alpha_{13}$ are,
$$
\alpha_{13} = \begin{cases} -0.10 \pm 0.16 \quad &\text{({\tt IMRPhenomPv2})} \\ 0.08 \pm 0.19 \quad &\text{({\tt SEOBNRv4P})} \end{cases}.
$$
The constraints on individual GWTC-2 events can be found in Ref.~\cite{shashank2022}.

\section{Simpson-Visser regular BH metric}
We also followed the same calculations for a regular BH metric coined by Simpson and Visser \cite{Simpson:2018tsi,mazza2021novel}. This metric has a regularization parameter $l$ which in our analysis we assume as a deformation and follow the analysis as in the previous sections. The metric reads,
\be
    \label{eq:sv_metric}
	d s^2 = -\left(1-\frac{2M}{\Delta}\right) d t^{2} + \left(1-\frac{2M}{\Delta}\right)^{-1}  d r^{2}
	+ \Delta^2 \left( d \theta^{2} + d \phi^{2} \right) \, ,
\ee
where $\Delta = \sqrt{r^2 + l^2}$. Here, $l > 0$ regularizes the singularity at the centre and different values of $l$ gives rise to different scenarios. $l<2M$ is a regular BH, $l=2M$ gives rise to a one-way wormhole and $l>2M$ forms a traversable two-way wormhole.
The calculations following $l$ as a deformation to Kerr leads,
\be \label{eq:mod_kep_sv}
    \Omega^2 = \frac{M}{r^3} \left[ 1 + \frac{3 M}{r} + \frac{9 M^2}{r^2} - \frac{3 M^2}{r^2}l^2 + \mathcal{O} \left(l^4, \frac{M^3}{r^3} \right) \right] \, . \nonumber\\
\ee
and
\be\label{eq:l-p4}
   l^2 = -\frac{1}{100} \varphi_{4} \delta \varphi_{4} \, .
\ee
The parameter $l$, similar to $\delta_1$ of KRZ metric and $\alpha_{13}$ of Johannsen metric enters at 2PN. The constraints for $l$ could not be combined as we constrained values of $l^2$, of which we have to reject the negative values so we provide there the stringest constraint we obtained with the event GW190707A,
$$
l/M = \begin{cases} 0.44^{+0.24} \quad &\text{({\tt IMRPhenomPv2})} \\ 0.44^{+0.28} \quad &\text{({\tt SEOBNRv4P})} \end{cases}.
$$
The constraints on rest of the GWTC-2 events can be found in Ref.~\cite{Riaz:2022rlx}.

\section{Conclusions}
\begin{figure}
    \centering
    \includegraphics[width=0.9\linewidth]{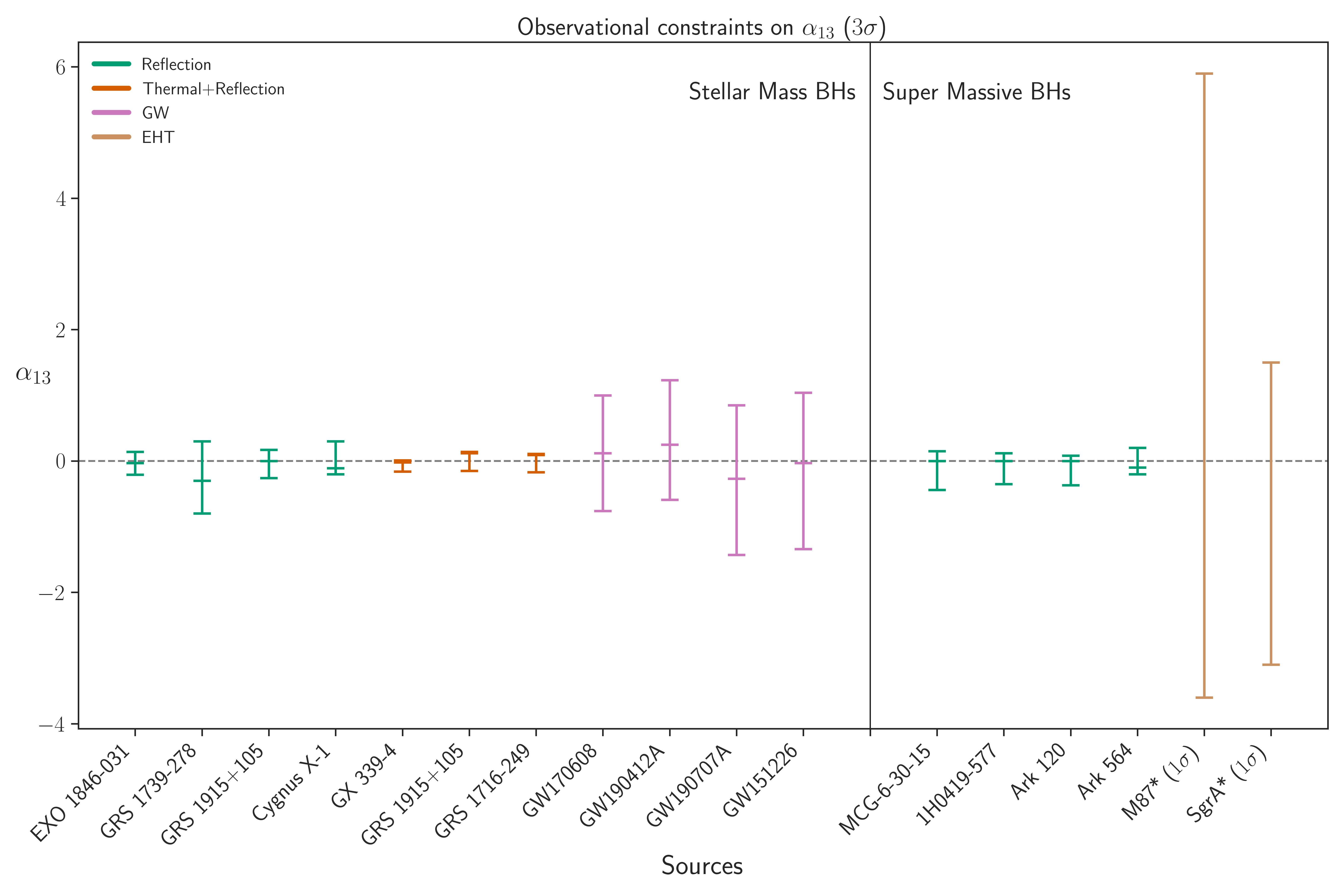}
    \caption{Constraints for $\alpha_{13}$ parameter of the Johannsen metric \cite{Johannsen:2013szh} across X-ray, radio and GW measurements. Data from Refs.~\cite{Bambi:2021chr,shashank2022,EventHorizonTelescope:2022xqj,psaltis2020gravitational}.}
    \label{fig:a13_multi}
\end{figure}

In the previous sections, we see the possibility of testing the Kerr nature of black holes by using the GW data against parametrized metrics. Such tests also act as agnostic tests of GR and also help in comparing the GR tests across multi-messenger observations as shown in Fig.~\ref{fig:a13_multi}.

%
%
\bibliographystyle{acm}
\bibliography{ref}
\end{document}